\documentclass{article}

\usepackage{moreverb,url}
\usepackage{amsmath,amssymb}
\usepackage{graphicx}
\usepackage[margin=1.5in]{geometry}
\usepackage[]{hyperref}

\date{}

\begin{document}

\title{The COVID-19 Infodemic: Twitter versus Facebook}

\author{Kai-Cheng Yang,\textsuperscript{1} Francesco Pierri,\textsuperscript{1, 2} Pik-Mai Hui,\textsuperscript{1} David Axelrod,\textsuperscript{1} \\ Christopher Torres-Lugo,\textsuperscript{1} John Bryden,\textsuperscript{1} and Filippo Menczer\textsuperscript{1}  \vspace{1em}\\
\textsuperscript{1}Observatory on Social Media, Indiana University, Bloomington, USA\\
\textsuperscript{2}Dipartimento di Elettronica, Informatica e Bioingegneria,\\Politecnico di Milano, Italy
}

\maketitle

\begin{abstract}
The global spread of the novel coronavirus is affected by the spread of related misinformation --- the so-called COVID-19 Infodemic --- that makes populations more vulnerable to the disease through resistance to mitigation efforts.  
Here we analyze the prevalence and diffusion of links to low-credibility content about the pandemic across two major social media platforms, Twitter and Facebook. We characterize cross-platform similarities and differences in popular sources, diffusion patterns, influencers, coordination, and automation. 
Comparing the two platforms, we find divergence among the prevalence of popular low-credibility sources and suspicious videos.
A minority of accounts and pages exert a strong influence on each platform. These misinformation ``superspreaders'' are often associated with the low-credibility sources and tend to be verified by the platforms.
On both platforms, there is evidence of coordinated sharing of Infodemic content. 
The overt nature of this manipulation points to the need for societal-level solutions in addition to  mitigation strategies within the platforms. However, we highlight limits imposed by inconsistent data-access policies on our capability to study harmful manipulations of information ecosystems.
\end{abstract}

\maketitle

\section{Introduction}

The impact of the COVID-19 pandemic has been felt globally, with almost 70 million detected cases and 1.5 million deaths as of December 2020 (\url{coronavirus.jhu.edu/map.html}). Epidemiological strategies to combat the virus require collective behavioral changes. To this end, it is important that people receive coherent and accurate information from media sources that they trust. Within this context, the spread of false narratives in our information environment can have acutely negative repercussions on public health and safety. For example, misinformation about masks greatly contributed to low adoption rates and increased disease transmission~\cite{lyu_community_2020}. The problem is not going away any time soon: false vaccine narratives~\cite{loomba2021measuring} will drive hesitancy, making it difficult to reach herd immunity and prevent future outbreaks.

It is concerning that many people believe, and many more have been exposed to, misinformation about the pandemic~\cite{mitchell_americans_2020,schaeffer_nearly_2020,nightingale_quantifying_2020,roozenbeek_susceptibility_2020}. The spread of this misinformation has been termed the \textit{Infodemic}~\cite{zarocostas2020fight}. Social media play a strong role in propagating misinformation because of peer-to-peer transmission~\cite{vosoughi2018spread}. There is also evidence that social media are manipulated~\cite{shao2018spread,stella2018bots} and used to spread COVID-19 misinformation~\cite{ferrara_misinformation_2020}. It is therefore important to better understand how users disseminate misinformation across social media networks.

As described in the literature reviewed in the next section, a limitation of most existing studies regarding misinformation spreading on social media is that they  focus on a single platform.
However, the modern information ecosystem consists of different platforms over which information propagates concurrently and in diverse ways.
Each platform can have different vulnerabilities~\cite{allcott_trends_2019}.
A key goal of the present work is to compare and contrast the extent to which the Infodemic has spread on Twitter and Facebook.

A second gap in our understanding of COVID-19 misinformation is in the patterns of diffusion within social media. It is important to understand how certain user accounts, or groups of accounts, can play a disproportionate role in amplifying the spread of misinformation. 
Inauthentic social media accounts, known as social bots and trolls, can play an important role in amplifying the spread of COVID-19 misinformation on Twitter~\cite{ferrara_what_2020}. The picture on Facebook is less clear, as there is little access to data that would enable a determination of social bot activity. It is, however, possible to look for evidence of manipulation in how multiple accounts can be coordinated with one another, potentially controlled by a single entity. For example, accounts may exhibit suspiciously similar sharing behaviors~\cite{pacheco2020uncovering}.

We extract website links from social media posts that include COVID-19 related keywords. We identify a link with low-credibility content in one of two ways. First, we follow the convention of classifying misinformation at the source rather than the article level~\cite{Lazer-fake-news-2018}. We do this by matching links to an independently-generated corpus of low-credibility website \textit{domains} (or \textit{sources}).
Second, in the case of links to YouTube, we label videos as suspicious if they have been banned by the site or are otherwise unavailable to the public. This enables us to quantify the prevalence of individual uploads likely to propagate COVID-19 misinformation and the different ways in which they are shared on Twitter and Facebook.

The main contributions of this study stem from exploring three sets of research questions:
\begin{enumerate}

\item What is the prevalence of low-credibility content on Twitter and Facebook? Are there similarities in how sources are shared over time? How does this activity compare to that of popular high-credibility sources? Are the same suspicious sources and YouTube videos shared in similar volumes across the two platforms? 

\item Is the sharing of misinformation concentrated around a few active accounts? 
Do a few influential accounts dominate the resharing of popular misinformation? 
What is the role of verified accounts and those associated with the low-credibility sources on the two platforms?

\item Is there evidence of inauthentic coordinated behavior in sharing low-credibility content?  Can we identify clusters of users, pages, or groups with suspiciously similar sharing patterns? Is low-credibility content amplified by Twitter bots more prevalent on Twitter as compared to Facebook?
\end{enumerate}

After systematically reviewing the literature regarding health misinformation on social media in the next section,  we describe the methodology and data employed in our analyses. The following three sections present results to answer the above research questions. Finally, we discuss the limitations of our analyses and implications of our findings for mitigation strategies.

\section{Literature review}

Concerns regarding online health-related misinformation existed before the advent of online social media. Studies mostly focused on evaluating the quality of information on the web~\cite{eysenbach2002empirical}, and
a new research field emerged, namely ``infodemiology,'' to assess health-related information on the Internet and address the gap between expert knowledge and public perception~\cite{eysenbach2002infodemiology}.

With the wide adoption of online social media, the information ecosystem has seen large changes. Peer-to-peer communication can greatly amplify fake or misleading messages by any individual~\cite{vosoughi2018spread}. 
Many studies reported on the presence of misinformation on social media during the time of epidemics such as Ebola~\cite{jin2014misinformation,pathak2015youtube,fung2016social,sell2020misinformation} and Zika~\cite{seltzer2017public,sharma2017zika,bora2018internet,wood2018propagating}.
Misinformation surrounding vaccines has been particularly persistent and is likely to reoccur whenever the topic comes into public focus~\cite{mahoney2015digital,bahk2016publicly,donzelli2018misinformation,panatto2018comprehensive,schmidt2018polarization,deverna2021covaxxy}.

These studies focused on specific social media platforms including Twitter~\cite{mahoney2015digital,wood2018propagating},  Facebook~\cite{sharma2017zika,schmidt2018polarization}, Instagram~\cite{seltzer2017public}, and YouTube~\cite{donzelli2018misinformation,bora2018internet}.
The most common approach was content-based analysis of sampled social media posts, images, and videos to gauge the topics of online discussions and estimate the prevalence of misinformation.
Unfortunately, the datasets analysed in these studies were usually small (at a scale of hundreds or thousands of items) due to difficulties in accessing and manually annotating large scale collections.

Unsurprisingly, the COVID-19 pandemic has inspired a new wave of health misinformation studies.
In addition to traditional approaches like qualitative analyses of social media content~\cite{pulido_covid-19_2020,al-rakhami_lies_2020,memon_characterizing_2020,kouzy2020coronavirus,li2020youtube,Dutta_YTcovid_2020} and survey studies~\cite{mitchell_americans_2020,nightingale_quantifying_2020,roozenbeek_susceptibility_2020}, quantitative studies on the prevalence of links to low-credibility websites at scale have gained popularity in light of the recent development of computational methods~\cite{broniatowski_covid-19_2020,cinelli_covid-19_2020,gallotti2020assessing,singh_first_2020,yang2020prevalence,singh2020understanding,guarino2021information}.

Many of these studies aimed to assess the prevalence of, and exposure to, COVID-19 misinformation on online social media~\cite{chou2018addressing}.
However, different approaches yielded disparate estimates of misinformation prevalence levels ranging from as little as 1\% to as much as 70\%.
These widely varying statistics indicate that different approaches to experimental design, including uneven access to data on different platforms and inconsistent definitions of misinformation, can generate inconclusive or misleading results.
In this study, we follow the reasoning from Gallotti et al. that it is better to clearly define a misinformation metric, and then use it in a comparative way to look at how misinformation varies over time or is influenced by other factors~\cite{gallotti2020assessing}.

We identify two gaps in the literature reviewed above.
First, it is still unclear how the spreading patterns can differ on different social networks since studies comparing multiple platforms are rare.
This might be due to the obstacles in accessing data from different sources simultaneously and the lack of a unified framework to compare very different services.
Second, our understanding of the role that different account groups play during the misinformation dissemination is very limited.
We hope to address these gaps in the present study.

\section{Methods}

In this section we describe in detail the methodology employed in our analyses, allowing other researchers to replicate our approach. The outline is as follows: we collect social media data from Twitter and Facebook using the same keywords list.
We then identify low- and high-credibility content from the tweets and posts automatically by tracking the URLs linking to the domains in a pre-defined list.
Finally, we identify suspicious YouTube videos by their availability status.

\subsection{Identification of low-credibility information}
\label{sec:low_credibility}

We focus on news articles linked in social media posts and identify those pertaining to low-credibility domains by matching the URLs to sources, following a corpus of literature~\cite{Lazer-fake-news-2018,shao2018spread,grinberg2019fake,pennycook2019fighting,bovet2019influence}.
We define our list of low-credibility domains based on information provided by the Media Bias/Fact Check website (MBFC, \url{mediabiasfactcheck.com}), an independent organization that reviews and rates the reliability of news sources.
We gather the sources labeled by MBFC as having a ``Very Low'' or ``Low'' factual-reporting level. We then add ``Questionable'' or ``Conspiracy-Pseudoscience'' sources and we leave out those with factual-reporting levels of ``Mostly-Factual,'' ``High,'' or ``Very High.''
We remark that although many websites exhibit specific political leanings, these do not affect inclusion in the list.
The list has 674 low-credibility domains~\cite{dataset}.

\subsection{High-credibility sources}

\begin{table}
\caption{
List of high-credibility sources.
}
\centering
\begin{tabular}{ll}
\hline
huffpost.com        &  newyorker.com \\
msnbc.com           &  newsweek.com  \\
cnn.com             &  nytimes.com  \\
economist.com       &  time.com \\
washingtonpost.com  &  reuters.com \\
apnews.com          &  npr.org \\
usatoday.com        &  wsj.com  \\
foxnews.com         &  marketwatch.com \\
nypost.com          &  dailycaller.com \\
theblaze.com        &  dailywire.com \\
\hline
cdc.gov             &  who.int \\
\hline
\end{tabular}
\label{tab:high_credibility}
\end{table}

As a benchmark for interpreting the prevalence of low-credibility content, we also curate a list of 20 more credible information sources. We start from the list provided in a recent Pew Research Center report~\cite{pew2014} and used in a few studies on online disinformation~\cite{Pierri2020epj,Pierri2020scirep}, and we select popular news outlets that cover the full U.S. political spectrum. These sources have a MBFC factual-reporting level of ``Mixed'' or higher. 
In addition, we include the websites of two organizations that acted as authoritative sources of COVID-19 related information, namely the Centers for Disease Control and Prevention (CDC) and World Health Organization (WHO).
For simplicity we refer to the full list in Table~\ref{tab:high_credibility} as high-credibility sources.

\subsection{Data collection}
\label{sec:data_collection}

We collect data related to COVID-19 from both Twitter and Facebook.
To provide a general and unbiased view of the discussion, we chose the following generic query terms: \texttt{coronavirus}, \texttt{covid} (to capture keywords like \texttt{covid19} and \texttt{covid-19}), and \texttt{sars} (to capture \texttt{sars-cov-2} and related variations).

\subsubsection{Twitter data.}

\begin{figure}
    \centering
    \includegraphics[width=0.6\columnwidth]{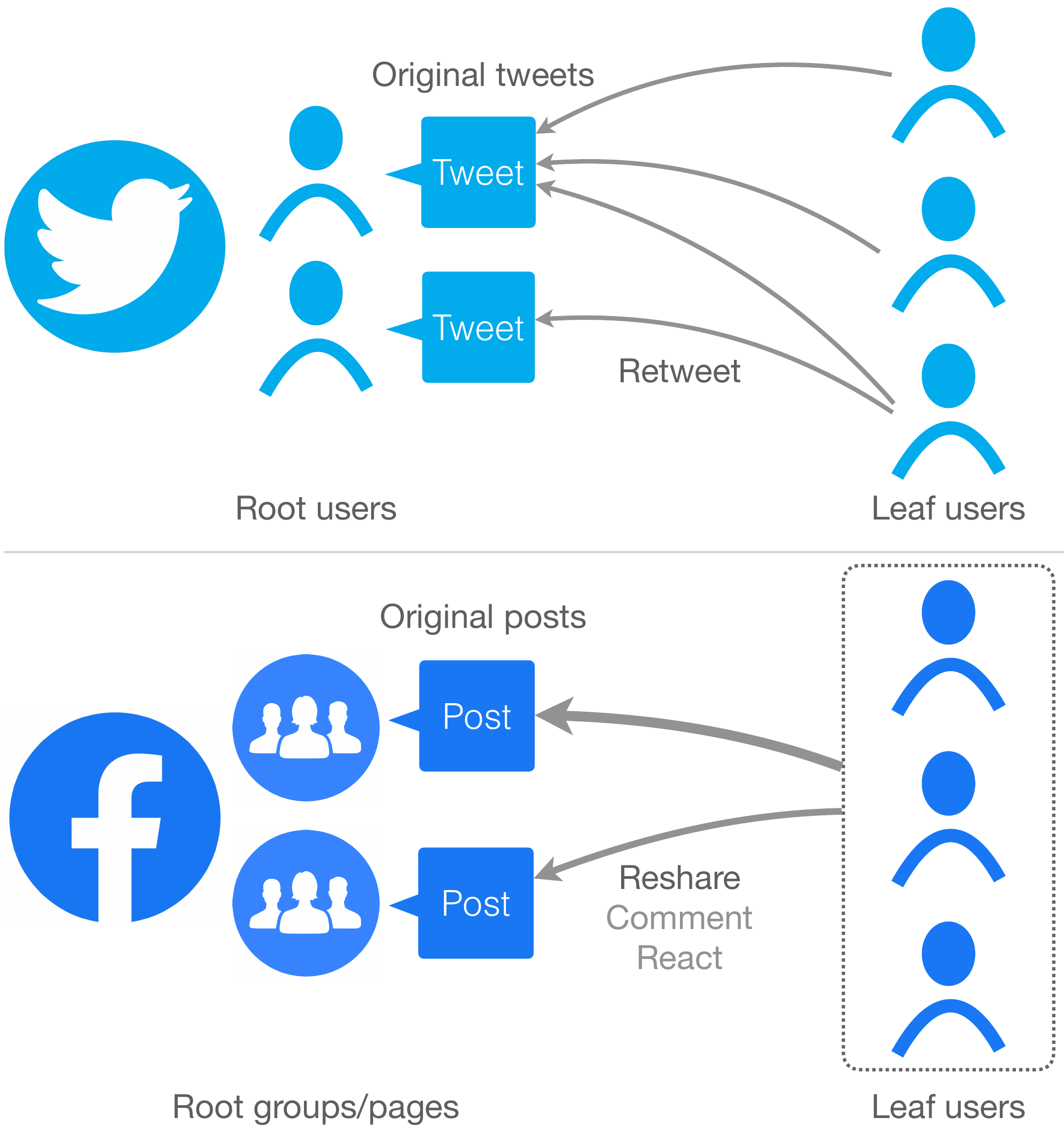}
    \caption{
    Structure of the data collected from Twitter and Facebook.
    On Twitter, we have the information about original tweets, retweets, and all the accounts involved.
    On Facebook, we have information about original posts and public groups/pages that posted them. For each post, we also have aggregate numbers of reshares, comments, and reactions, with no information about the users responsible for those interactions.
    }
    \label{fig:tw_fb_illustration}
\end{figure}

Our Twitter data was collected using an API from the Observatory on Social Media~\cite{davis2016osome}, which allows to search tweets from the Decahose, a 10\% random sample of public tweets. 
We searched for English tweets containing the keywords between Jan. 1 and Oct. 31, 2020, resulting in over 53M tweets posted by about 12M users.
Note that since the Decahose samples tweets and not users, the sample of users in our Twitter dataset is biased toward more active users.

Our collection contains two types of tweets, namely original tweets and retweets. 
The content of original tweets is published by users directly, while retweets are generally used to endorse/amplify original tweets by others (no quoted tweets are included). 
We refer to authors of original tweets as ``root'' users, and to authors of retweets as ``leaf'' users (see Fig.~\ref{fig:tw_fb_illustration}). 

\subsubsection{Facebook data.}

We used the \textit{posts/search} endpoint of the CrowdTangle API \cite{crowdtangle} to collect data from Facebook.
We filtered the entire set of English posts published by public pages and groups in the period from Jan. 1 to Oct. 31, 2020 using the above list of keywords, resulting in over 37M posts by over 140k public pages/groups.

Our Facebook data collection is limited by the coverage of pages and groups in CrowdTangle, a public tool owned and operated by Facebook. CrowdTangle includes over 6M Facebook pages and groups: all those with at least 100k followers/members, U.S. based public groups with at least 2k members, and a very small subset of verified profiles that can be followed like public pages. We include these public accounts among pages and groups. In addition, some pages and groups with fewer followers and members are also included by CrowdTangle upon request from users. This might bias the dataset in ways that are hard to gauge. For example, requests from researchers interested in monitoring low-credibility pages and groups might lead to over-representation of such content.

\begin{figure}
    \centering
    \includegraphics[width=0.6\columnwidth]{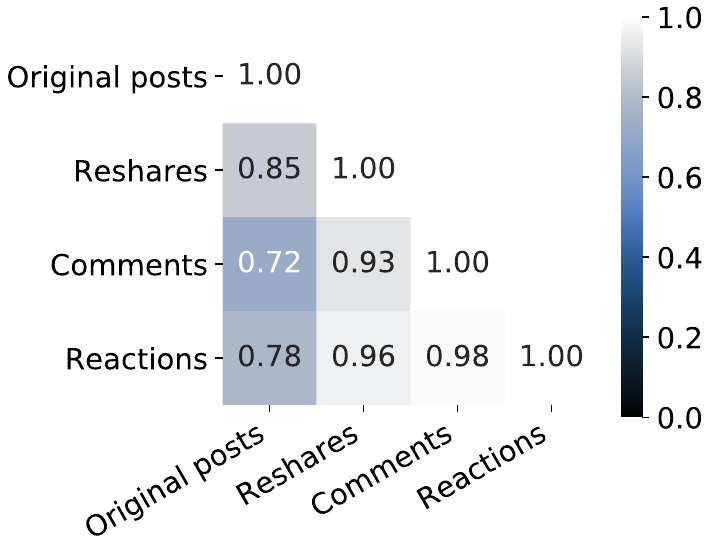}
    \caption{
    Pearson correlation coefficients between Facebook metrics aggregated at the domain level for low-credibility domains.
    A reaction can be a ``like,'' ``love,'' ``wow,'' ``haha,'' ``sad,'' ``angry,'' or ``care.''
    All correlations are significant ($p<0.01$).
    }
    \label{fig:fb_correlation}
\end{figure}

As shown in Fig.~\ref{fig:tw_fb_illustration}, the collected data contains information about original Facebook posts and the pages/groups that published these posts. 
For each post, we also have access to \textit{aggregate} statistics such as the number of reshares, comments, and reactions (e.g., ``likes'') by Facebook users. 
The numbers of comments and reactions are highly correlated with reshares (Fig.~\ref{fig:fb_correlation}), so we focus on reshares in this study.

Similarly to Twitter, Facebook pages and groups that publish posts are referred to as ``roots'' and users who reshare them are ``leaves.''
However, in contrast to Twitter, we don't have access to any information about leaf users on Facebook.
We refer generically to Twitter users and Facebook pages and groups as ``accounts.'' 

To compare Facebook and Twitter in a meaningful way, we compare root users with root pages/groups, original tweets with original posts, and retweet counts with reshare counts. We define \textit{prevalence} as the sum of original tweets and retweets on Twitter, and as the sum of original posts and reshares on Facebook.

\subsubsection{YouTube data.}

We observed a high prevalence of links pointing to \texttt{youtube.com} on both platforms --- over 64k videos on Twitter and 204k on Facebook. Therefore, we also provide an analysis of popular videos published on Facebook and Twitter. Specifically, we focus on popular YouTube videos that are likely to contain low-credibility content. 
An approach analogous to the way we label links to websites would be to identify sources that upload low-credibility videos and then label every video from those sources as misinformation. However, this approach is infeasible because the list of YouTube channels would be huge and fluid.  
To circumvent this difficulty, we use removal of videos by YouTube as a proxy to label low-credibility content. 
We additionally consider private videos to be suspicious, since this can be used as a tactic to evade the platform's sanctions when violating terms of service. 

To identify the most popular and suspicious YouTube content, we first select the 16,669 videos shared at least once on both platforms. 
We then query the YouTube API \textit{Videos:list} endpoint to collect their metadata and focus on the 1,828 (11\%) videos that had been removed or made private.
To validate this approach for identifying low credibility YouTube content, we follow a two-step manual inspection process for a sample of about 3\% of the unavailable videos, comprising a mix of randomly selected and popular ones.
We first search for the deleted video IDs in other YouTube videos and web pages.
When these references contain the deleted videos' titles, we search for these titles on \url{bitchute.com} to find copies of the original videos.
This process allows us to identify the narratives of 40 deleted videos, 90\% of which contain misinformation.
A similar approach was also adopted by \cite{knuutila2020covid} in their recent study of COVID-19 misinformation on YouTube.

\subsubsection{Ethical considerations.}

Our studies of public Twitter and Facebook data have been granted exemption from IRB review (Indiana University protocols 1102004860 and 10702, respectively). 
The data collection and analysis are also in compliance with the terms of service of the corresponding social media platforms.

\subsection{Link extraction}

\begin{table}
\caption{
List of URL shortening services.
}
\centering
\begin{tabular}{llll}
\hline
     bit.ly &       dlvr.it &  liicr.nl &  tinyurl.com \\
     goo.gl &        ift.tt &     ow.ly &       fxn.ws \\
    buff.ly &       back.ly &   amzn.to &      nyti.ms \\
     nyp.st &  dailysign.al &      j.mp &      wapo.st \\
    reut.rs &     drudge.tw &   shar.es &      sumo.ly \\
 rebrand.ly &    covfefe.bz &   trib.al &      yhoo.it \\
       t.co &        shr.lc &     po.st &       dld.bz \\
  bitly.com &      crfrm.us &   flip.it &        mf.tt \\
      wp.me &       voat.co &   zurl.co &        fw.to \\
     mol.im &       read.bi &   disq.us &    tmsnrt.rs \\
    usat.ly &        aje.io &     sc.mp &       gop.cm \\
    crwd.fr &        zpr.io &    scq.io &      trib.in \\
     owl.li &               &           &              \\
\hline
\end{tabular}
\label{tab:url_shortening_service}
\end{table}

Estimating the prevalence of low-credibility information requires matching URLs, extracted from tweets and Facebook metadata, against our lists of low- and high-credibility websites.
As shortened links are very common, we also identified 49 link shortening services that appear at least 50 times in our datasets (Table~\ref{tab:url_shortening_service}) and expanded shortened URLs referring to these services through HTTP requests to obtain the actual domains. 
We finally match the extracted and expanded links against the lists of low- and high-credibility domains.
A breakdown of matched posts/tweets is shown in Table~\ref{table:fb_tw_stats}. For low-credibility content, the ratio of retweets to tweets is 2.7:1, while the ratio of reshares to posts is 68:1.
This large discrepancy is due to various factors: the difference in traffic on the two platforms, the fact that we only have a 10\% sample of tweets, and the bias toward popular pages and groups on Facebook. 

\begin{table}
\centering
\caption{Breakdown of Facebook and Twitter posts/tweets matched to low- and high-credibility domains.}
\begin{tabular}{lrr}
\hline
                   & Low-credibility & High-credibility \\
\hline
\textbf{Facebook}  &                 &             \\
Original posts     &     303,119     &   1,194,634 \\
Reshares           &  20,462,035     &  98,415,973 \\
\hline
\textbf{Twitter}   &                 &             \\
Original tweets    &    245,620      &   734,409          \\
Retweets           & 653,415         &  2,184,050 \\
\hline
\end{tabular}
\label{table:fb_tw_stats}
\end{table}

\section{Infodemic prevalence}

In this section, we provide results about the prevalence of links to low-credibility domains on the two platforms. As described in the Methods section, we sum tweets and retweets for Twitter, and original posts and reshares for Facebook.
Note that deleted content is not included in our data. Therefore, our estimations should be considered as lower bounds for the prevalence of low-credibility information on both platforms.

\subsection{Prevalence trends}

\begin{figure}
    \centering
    \includegraphics[width=0.5\columnwidth]{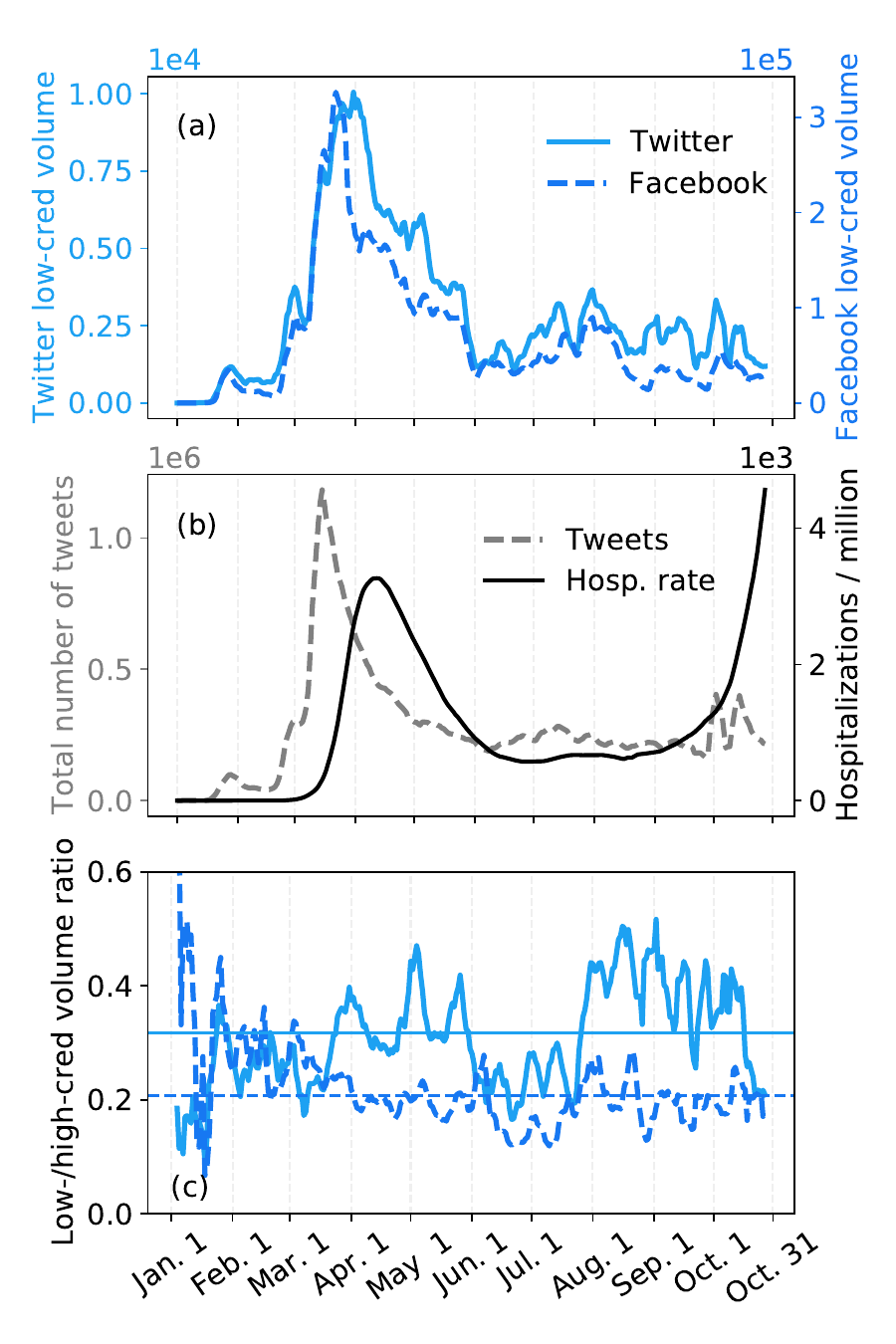}
    \caption{Infodemic content surge on both platforms around the COVID-19 pandemic waves, from Jan. 1 to Oct. 31, 2020. 
    All curves are smoothed via 7-day moving averages.
    \textbf{(a)}~Daily volume of posts/tweets linking to low-credibility domains on Twitter and Facebook.
    Left and right axes have different scales and correspond to Twitter and Facebook, respectively.
    \textbf{(b)}~Overall daily volume of pandemic-related tweets and worldwide COVID-19 hospitalization rates (data source: Johns Hopkins University).  
    \textbf{(c)}~Daily ratio of volume of low-credibility links to volume of high-credibility links on Twitter and Facebook. 
    The noise fluctuations in early January are due to low volume. The horizontal lines indicate averages across the period starting Feb. 1.
    }
    \label{fig:timeseries}
\end{figure}

We plot the daily prevalence of links to low-credibility sources on Twitter and Facebook in Fig.~\ref{fig:timeseries}(a).
The two time series are strongly correlated (Pearson $r=0.87$, $p < 0.01$).
They both experience a drastic growth during March, when the number of COVID-19 cases was growing worldwide. Towards summer, the prevalence of low-credibility information decreases to a relatively low level and then becomes more stable.

To analyze the Infodemic surge with respect to the pandemic's development and public awareness,  Fig.~\ref{fig:timeseries}(b) shows the worldwide hospitalization rate and the overall volume of tweets in our collection.  The Infodemic surge roughly coincides with the general attention given to the pandemic, captured by the overall Twitter volume. The peak in hospitalizations trails by a few weeks. 
A similar delay was recently reported between peaks of exposure to Infodemic tweets and of COVID-19 cases in different countries~\cite{gallotti2020assessing}. 
This plot suggests that the delay is related to general attention toward the pandemic rather than specifically toward misinformation. 

To further explore whether the decrease in low-credibility information is organic or due to platform interventions, we also compare the prevalence of low-credibility content to that of links to credible sources. As shown in Fig.~\ref{fig:timeseries}(c), the ratios are relatively stable across the observation period.
These results suggest that the prevalence of low-credibility content is mostly driven by the public attention to the pandemic in general, which progressively decreases after the initial outbreak.
We finally observe that Twitter exhibits a higher ratio of low-credibility information than Facebook (32\% vs.~21\% on average).

\subsection{Prevalence of specific domains}

\begin{figure*}
    \centering
    \includegraphics[width=\textwidth]{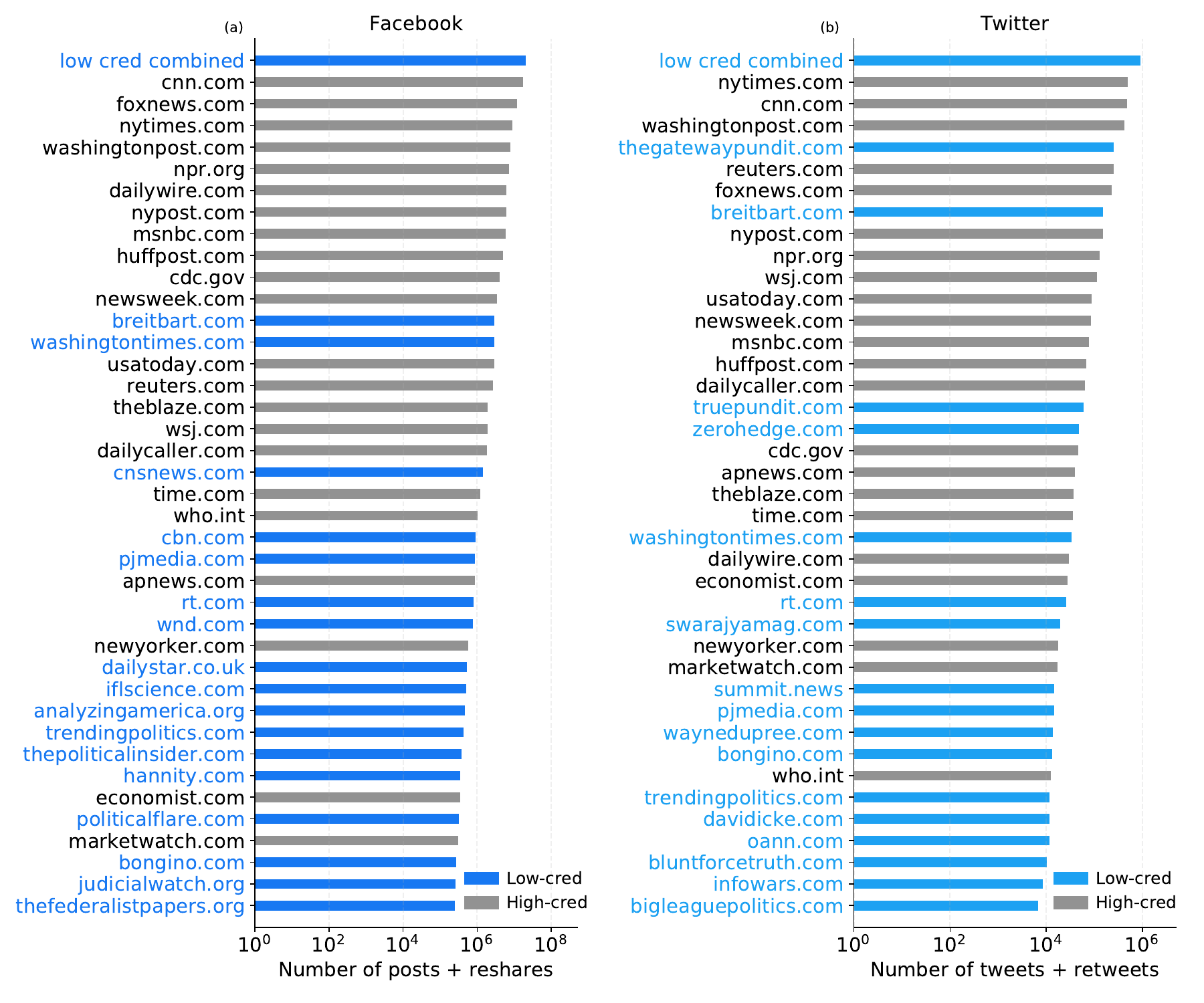}
    \caption{
    Total prevalence of links to low- and high-credibility domains on both \textbf{(a)} Facebook and \textbf{(b)} Twitter. 
    Due to space limitation, we only show the 40 most frequent domains on the two platforms.
    The high-credibility domains are all within the top 40.
    We also show low-credibility information as a whole (cf. ``low cred combined'').
    }
    \label{fig:prevalence}
\end{figure*}

We use the high-credibility domains as a benchmark to assess the prevalence of low-credibility domains on each platform. As shown in Fig.~\ref{fig:prevalence}, we notice that the low-credibility sources exhibit disparate levels of prevalence. 
Low-credibility content as a whole reaches considerable volume on both platforms, with prevalence surpassing every single high-credibility domain considered in this study. On the other hand, low-credibility domains generally exhibit much lower prevalence compared to high-credibility ones (with a few exceptions, notably \texttt{thegatewaypundit.com} and \texttt{breitbart.com}).

\subsection{Source popularity comparison}

\begin{figure*}
    \centering
    \includegraphics[width=\textwidth]{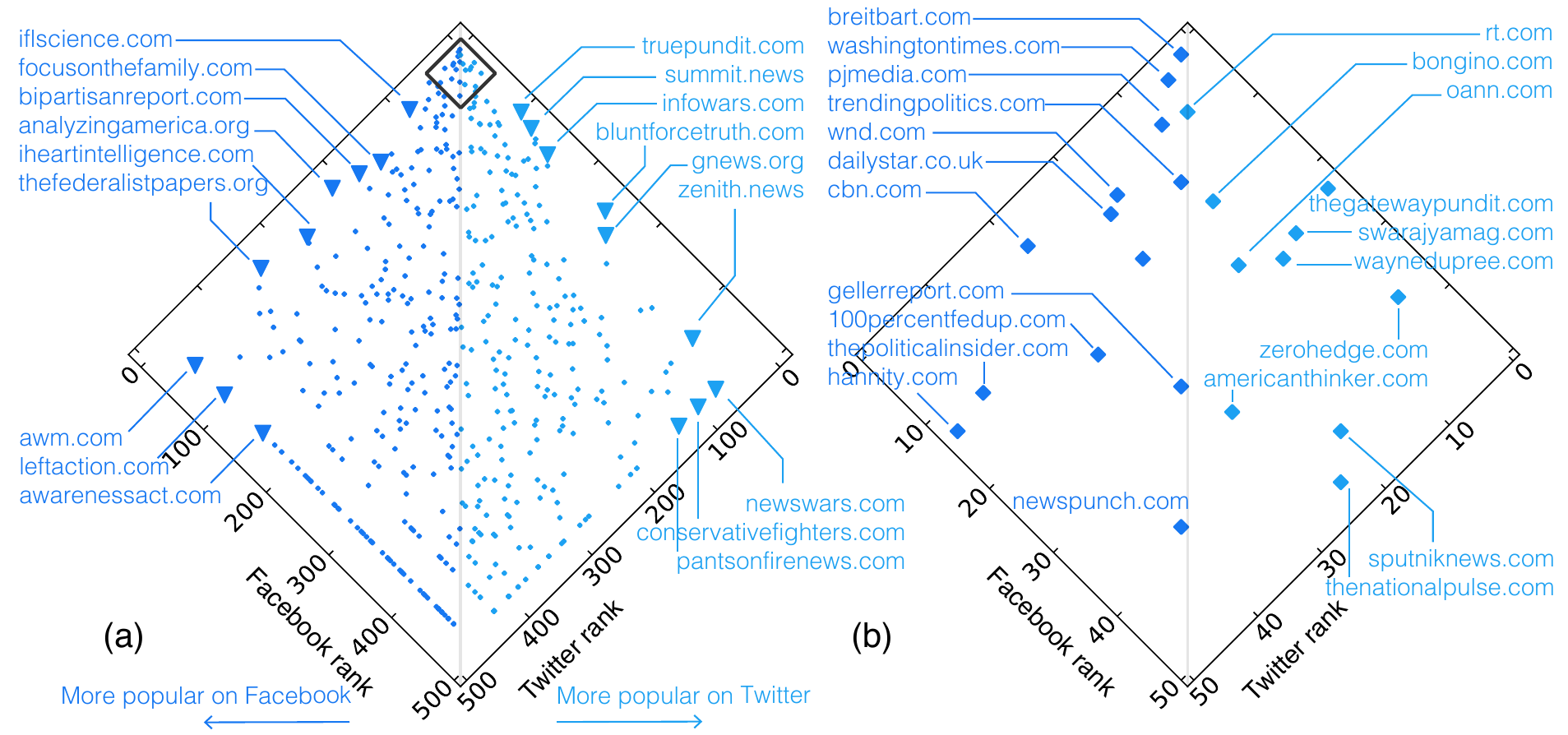}
    \caption{
    \textbf{(a)} Rank comparison of low-credibility sources on Facebook and Twitter.
    Each dot in the figure represents a low-credibility domain.
    The most popular domain ranks first.
    Domains close to the vertical line have similar ranks on the two platforms.
    Domains close to the edges are much more popular on one platform or the other.
    We annotate a few selected domains that exhibit high rank discrepancy.
    \textbf{(b)} A zoom-in on the sources ranked among the top 50 on both platforms (highlighted square in (a)).
    }
    \label{fig:rank_compare}
\end{figure*}

As shown in Fig.~\ref{fig:prevalence}, we observe that low-credibility websites may have different prevalence on the two platforms.
To further contrast their prevalence levels on Twitter and Facebook, we measure the popularity of websites on each platform by ranking them by prevalence, and then compare the resulting ranks in Fig.~\ref{fig:rank_compare}.
The ranks on the two platforms are not strongly correlated (Spearman $r=0.57$, $p < 0.01$). A few domains are much more popular or only appear on one of the platforms (see annotations in Fig.~\ref{fig:rank_compare}(a)).
We also show the domains that are very popular on both platforms in Fig.~\ref{fig:rank_compare}(b). They are dominated by right-wing and state sources, such as \texttt{breitbart.com}, \texttt{washingtontimes.com},  \texttt{thegatewaypundit.com}, \texttt{oann.com}, and \texttt{rt.com}. 

\subsection{YouTube Infodemic content}

Thus far, we examined the prevalence of links to low-credibility web page sources. However, a significant portion of the links shared on Twitter and Facebook point to YouTube videos, which can also carry COVID-19 misinformation. Previous work has shown that bad actors utilize YouTube in this manner for their campaigns~\cite{wilson2020cross}. 
Specifically, anti-scientific narratives on YouTube about vaccines, Idiopathic Pulmonary Fibrosis, and the COVID-19 pandemic have been documented~\cite{donzelli2018misinformation,Goobie_idiopathic_2019,Dutta_YTcovid_2020,knuutila2020covid}. 

\begin{figure*}
    \centering
    \includegraphics[width=\textwidth]{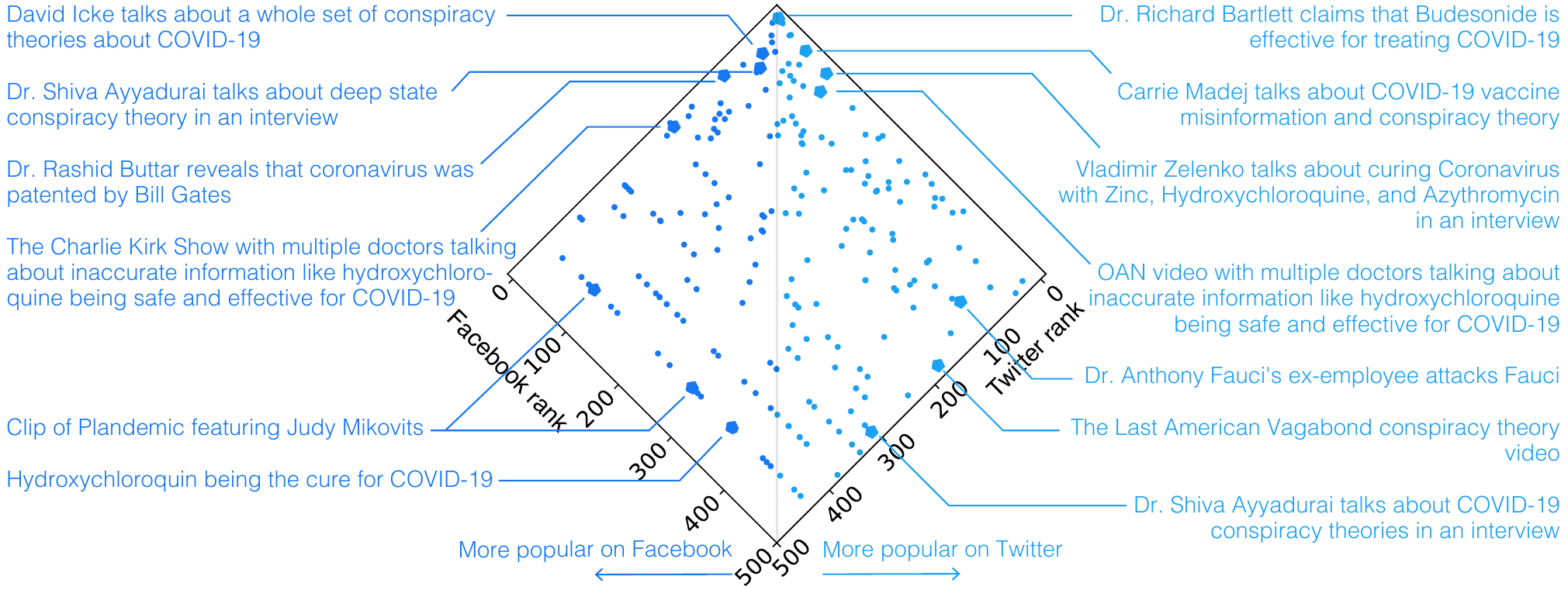}
    \caption{
    Rank comparison of suspicious YouTube videos within the top 500 on both Facebook and Twitter.
    The most popular video ranks first.
    Each dot in the figure represents a suspicious video.
    Videos close to the vertical line have similar ranks on both platforms.
    Videos close to the edges are more popular on one platform or the other.
    We annotated a few selected videos with their narratives extracted from their copies on \url{bitchute.com} or other web pages.
    }
    \label{fig:rank_corr_yt}
\end{figure*}

To measure the prevalence of Infodemic content introduced from YouTube, we consider the unavailability (deletion or private status) of videos as an indicator of suspicious content, as explained in the Methods section. 
Fig.~\ref{fig:rank_corr_yt} compares the prevalence rankings on Twitter and Facebook for unavailable videos ranked within the top 500 on both platforms. 
These videos are linked between 6 and 980 times on Twitter and between 39 and 64,257 times on Facebook. 
While we cannot apply standard rank correlation measures due to the exclusion of low-prevalence videos, we do not observe a correlation in the cross-platform popularity of suspicious content from a qualitative inspection of the figure. 
A caveat to this analysis is that the same video content (sometimes re-edited) can be recycled within many video uploads, each having a unique video ID. Some of these videos are promptly removed while others are not. Therefore, the lack of correlation could partly be driven by YouTube's efforts to remove Infodemic content in conjunction with attempts by uploaders to counter those efforts~\cite{knuutila2020covid}.

\begin{figure}
    \centering
    \includegraphics[width=0.6\columnwidth]{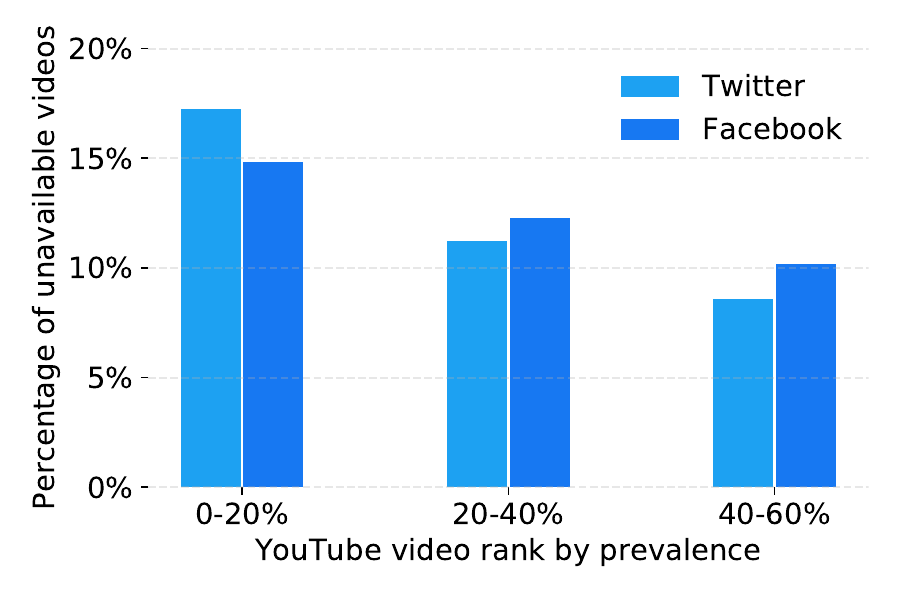}
    \caption{
    Percentages of suspicious YouTube videos against their percent rank among all videos linked from pandemic-related tweets/posts on both Twitter and Facebook.
    }
    \label{fig:video_percent_rank}
\end{figure}

Having looked at the prevalence of suspicious content from YouTube, we wish to explore the question from another angle: are videos that are popular on Facebook or Twitter more likely to be flagged as suspicious? Fig.~\ref{fig:video_percent_rank} shows this to be the case on both platforms: a larger portion of videos with higher prevalence are unavailable, but the trend is stronger on Twitter than on Facebook. The overall trend suggests that YouTube may have a bias toward moderating videos that attract more attention.
This may be a function of the fact that an Infodemic video that is spreading virally on Twitter/Facebook may receive more abuse reports through YouTube's reporting mechanism. The fact that this trend is greater on Twitter may be explained by the differences between each platform's demographics. Survey data cited in the Discussion section shows that Twitter users are younger and more educated; it is therefore plausible that the average Twitter user is more likely to report unreliable content.

\section{Infodemic spreaders}

Links to low-credibility sources are published on social media through original tweets and posts first, then retweeted and reshared by leaf users.
In this section, we study this dissemination process on Twitter and Facebook with a focus on the top spreaders, or ``superspreaders'': those disproportionately responsible for the distribution of Infodemic content.

\subsection{Concentration of influence}

We wish to measure whether low-credibility content originates from a wide range of users, or can be attributed to a few influential actors. 
For example, a source published 100 times could owe its popularity to 100 distinct users, or to a single root  whose post is republished by 99 leaf users. 
To quantify the concentration of original posts/tweets or reshares/retweets for a source $s$, we use the inverse normalized entropy~\cite{Nikolov2018biases}, defined as:
\begin{equation}
    \mathcal{C}_s = 1 + \sum_{r=1}^{N_s} \frac{P_r(s) \log{P_r(s)}}{\log{N_s}}\text{,}\nonumber
\end{equation}
where $r$ represents a root user/group/page linking to source $s$, $P_r(s)$ stands for the fraction of posts/tweets or reshares/retweets linking to $s$ and associated with $r$, and $N_s$ is the total number of roots linking to $s$. 
Entropy measures how evenly quantities of content are distributed across roots; it is normalized to account for the varying numbers of roots in different cases. Its inverse captures concentration, and is defined in the unit interval.
It is maximized at 1 when the content originates from a single root user/group/page, and minimized at 0 when each root makes an equal contribution. 
We set $\mathcal{C}_s=1$ when $N_s=1$.

\begin{figure*}
    \centering
    \includegraphics[width=\textwidth]{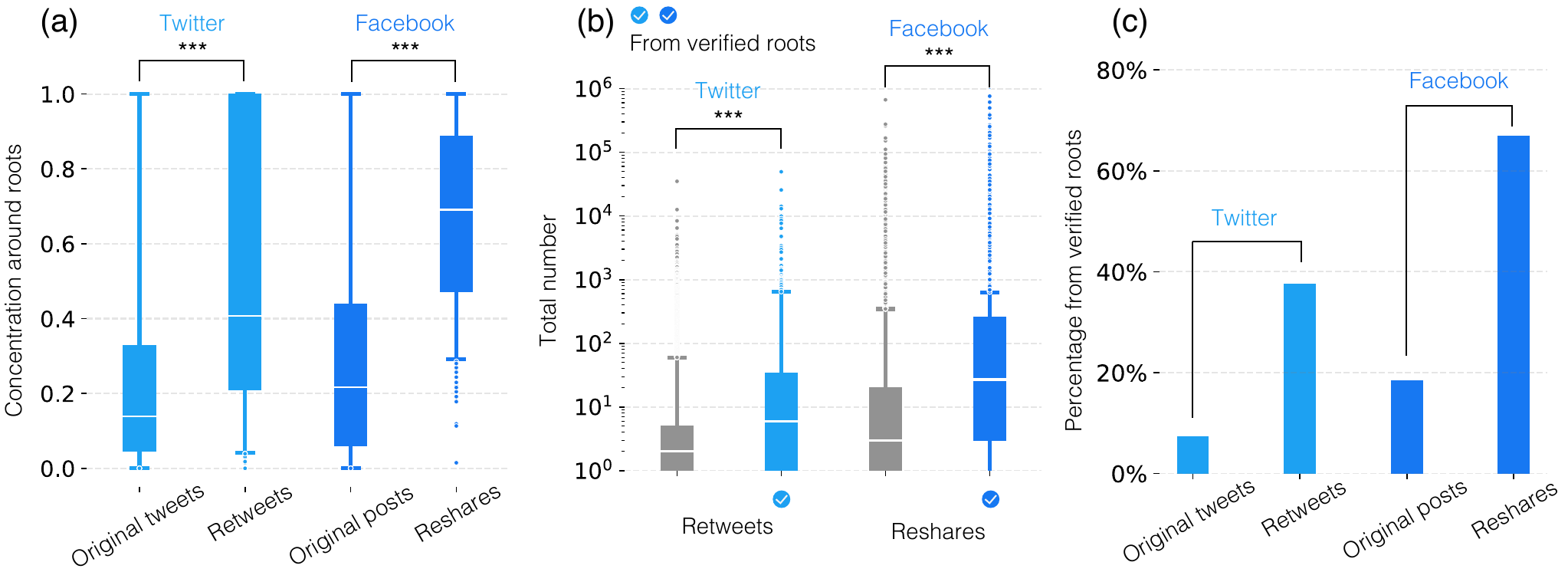}
    \caption{
    Evidence of Infodemic superspreaders. 
    Boxplots show the median (white line), 25th--75th percentiles (boxes), 5th--95th percentiles (whiskers), and outliers (dots).
    Significance of statistical tests is indicated by *** ($p<0.001$).
    \textbf{(a)} Distributions of the concentration of original tweets, retweets, original posts, and reshares linking to low-credibility domains around root accounts.
    Each domain corresponds to one observation.
    \textbf{(b)} Distributions of the total number of retweets and reshares of low-credibility content posted by verified and unverified accounts. 
    Each account corresponds to one observation.
    \textbf{(c)} Fractions of original tweets, retweets, original posts, and reshares by verified accounts. 
    }
    \label{fig:concentration_verified}
\end{figure*}

Let us gauge the \textit{concentration of activity} around root accounts through their numbers of original tweets/posts for each source.
Similarly, we calculate the \textit{concentration of popularity} around the root accounts using their numbers of retweets/reshares for each source. 
We show the distributions of these concentration variables in Fig.~\ref{fig:concentration_verified}(a).
On both platforms, we find that popularity is significantly more concentrated around root accounts compared to their activity ($p<0.001$ for paired sample t-tests).
This suggests the existence of superspreaders: despite the diversity of root accounts publishing links to low-credibility content on both platforms, only messages from a small group of influential accounts are shared extensively.

\subsection{Who are the Infodemic superspreaders?}

Both Twitter and Facebook provide verification of accounts and embed such information in the metadata. Although the verification processes differ, we wish to explore the hypothesis that verified accounts on either platform play an important role as top spreaders of low-credibility content. 
Fig.~\ref{fig:concentration_verified}(b) compares the popularity of verified accounts to unverified ones on a per-account basis. 
We find that verified accounts tend to receive a significantly higher number of  retweets/reshares on both platforms ($p<0.001$ for Mann-Whitney U-tests).

We further compute the proportion of original tweets/posts and retweets/reshares that correspond to verified accounts on both platforms.
Verified accounts are a small minority compared to unverified ones, i.e., 1.9\% on Twitter and 4.5\% on Facebook, among root accounts involved in publishing low-credibility content.
Despite this, Fig.~\ref{fig:concentration_verified}(c) shows that verified accounts yield almost 40\% of low-credibility retweets on Twitter and almost 70\% of reshares on Facebook.

These results suggest that verified accounts play an outsize role in the spread of Infodemic content. Are superspreaders all verified? To answer this question, let us analyse superspreader accounts separately for each low-credibility source. We extract the top user/page/group (i.e., the account with most retweets/reshares) for each source, and find that 19\% and 21\% of them are verified on Twitter and Facebook, respectively. While these values are much higher than the percentages of verified accounts among all roots, they show that not all superspreaders are verified.

\begin{table}
\caption{
Official social media handles for the 23 top low-credibility sources from Fig.~\ref{fig:rank_compare}(b).
Accounts with a checkmark (\checkmark) are verified.
Accounts with an asterisk (*) are the top spreaders for the corresponding domains.
Accounts with a dagger (\dag) were suspended as of February 2021.
}
\centering
\resizebox{\columnwidth}{!}{
\begin{tabular}{rll}
\hline
Domain                & Twitter handle                  & Facebook page/group  \\					
\hline
thegatewaypundit.com  & @gatewaypundit \checkmark * \dag& @gatewaypundit \checkmark *\\
breitbart.com         & @BreitbartNews \checkmark *     & @Breitbart \checkmark *\\
zerohedge.com         & @zerohedge *                    & @ZeroHedge *\\
washingtontimes.com   & @WashTimes \checkmark *         & @TheWashingtonTimes \checkmark \\
rt.com                & @RT\_com	\checkmark *        & @RTnews \checkmark *\\
swarajyamag.com       & @SwarajyaMag \checkmark *       & @swarajyamag \checkmark *\\
pjmedia.com           & @PJMedia\_com *                 & @PJMedia \checkmark *\\
waynedupree.com       & @WayneDupreeShow \checkmark *   & @WayneDupreeShow \checkmark *\\
bongino.com           & @dbongino \checkmark *          & @dan.bongino \checkmark *\\
trendingpolitics.com  & --                              & @trendingpoliticsdotcom \\
oann.com              & @OANN  \checkmark *             & @OneAmericaNewsNetwork \checkmark *\\				
wnd.com               & @worldnetdaily \checkmark *     & @WNDNews *\\
sputniknews.com	      & @SputnikInt \checkmark *        & @SputnikNews \checkmark *\\
dailystar.co.uk       & @dailystar \checkmark *         & @thedailystar \checkmark *\\
politicalflare.com    & @nicolejames                    & @nicolejameswriter \\
thenationalpulse.com  & @RaheemKassam \checkmark *      & @thenationalpulse *\\
americanthinker.com   & @AmericanThinker                & @AmericanThinker *\\
gellerreport.com      & @PamelaGeller \checkmark \dag   & @pamelageller \checkmark \\
cbn.com               & @CBNOnline \checkmark           & @cbnonline \checkmark \\																			
100percentfedup.com   & @100PercFEDUP *                 & @100PercentFEDUp *\\
newspunch.com         & --	                            & @thepeoplesvoicetv *\\
thepoliticalinsider.com & @TPInsidr                     & @ThePoliticalInsider \checkmark *\\
hannity.com           & @seanhannity \checkmark *       & @SeanHannity \checkmark *\\
\hline
\end{tabular}
}
\label{tab:top_low_cred_domains}
\end{table}

Who are the top spreaders of Infodemic content?  Table~\ref{tab:top_low_cred_domains} answers this question for the 23 top low-credibility sources in Fig.~\ref{fig:rank_compare}(b).  
We find that the top spreader for each source tends to be the corresponding official account.
For instance, about 20\% of the retweets containing links to \texttt{thegatewaypundit.com} pertain to \texttt{@gatewaypundit}, the official handle associated with \textit{The Gateway Pundit} website, on Twitter. (The \texttt{@gatewaypundit} account was suspended by Twitter in February 2021.) 
The remaining retweets have 10,410 different root users. 
Similarly, on Facebook, among all 2,821 pages/groups that post links to \texttt{thegatewaypundit.com}, the official page \texttt{@gatewaypundit} accounts for 68\% of the reshares.
We observe in Table~\ref{tab:top_low_cred_domains} that most of the top low-credibility sources have official accounts on both Twitter and Facebook, which tend to be verified (
71.4\% on Twitter and
65.2\% on Facebook).
They are also the top spreaders of those domains in 16 out of 21 cases (76.2\%) on Twitter and 18 out of 23 (78.3\%) on Facebook.

\section{Infodemic manipulation} 

Here we consider two types of inauthentic behaviors that can be used to spread and amplify COVID-19 misinformation: coordinated networks and automated accounts.

\subsection{Coordinated amplification of low-credibility content}

Social media accounts can act in a coordinated fashion (possibly controlled by a single entity) to increase influence and evade detection~\cite{nizzoli2020coordinated,sharma2020identifying,pacheco2020uncovering}.
We apply the framework proposed by \cite{pacheco2020uncovering} to identify coordinated efforts in promoting low-credibility information, both on Twitter and Facebook.

The idea is to build a network of accounts where the weights of edges represent how often two accounts link to the same domains. A high weight on an edge means that there is an unusually high number of domains shared by the two accounts.
We first construct a bipartite graph between accounts and low-credibility domains linked in tweets/posts.
The edges of the bipartite graph are weighted using TF-IDF~\cite{jones1972statistical} to discount the contributions of popular sources.
Each account is therefore represented as a TF-IDF vector of domains.
A projected co-domain network is finally constructed, with edges weighted by the cosine similarity between the account vectors.

We apply two filters to focus on active accounts and highly similar pairs. On Twitter, the users must have at least 10 tweets containing low-credibility links, and we retain edges with similarity above 0.99. On Facebook, the pages/groups must have at least 5 posts containing links, and we retain edges with similarity above 0.95. These thresholds are selected by manually inspecting the outputs.

\begin{figure*}
    \centering
    \includegraphics[width=0.8\textwidth]{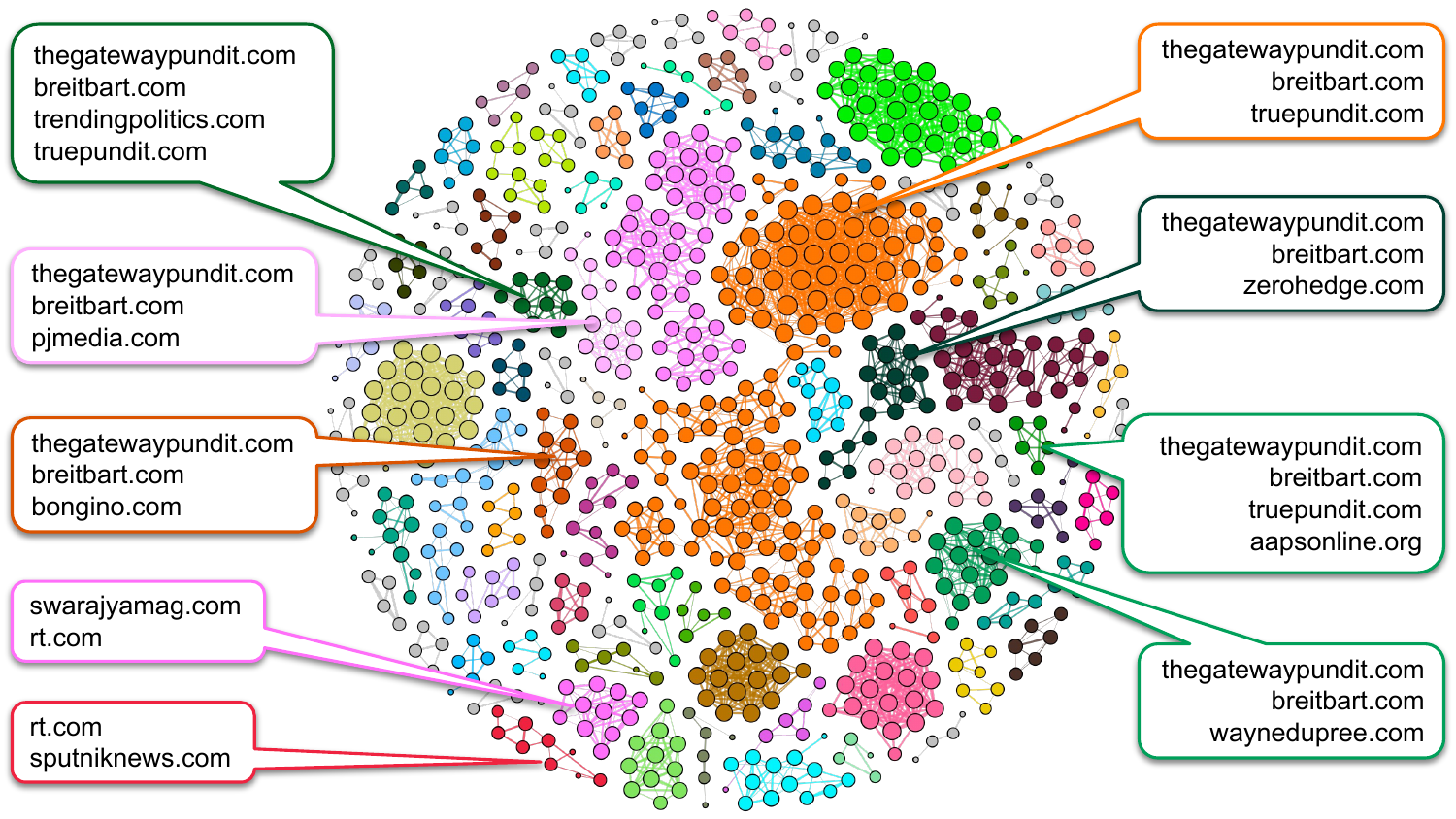}
    \includegraphics[width=0.8\textwidth]{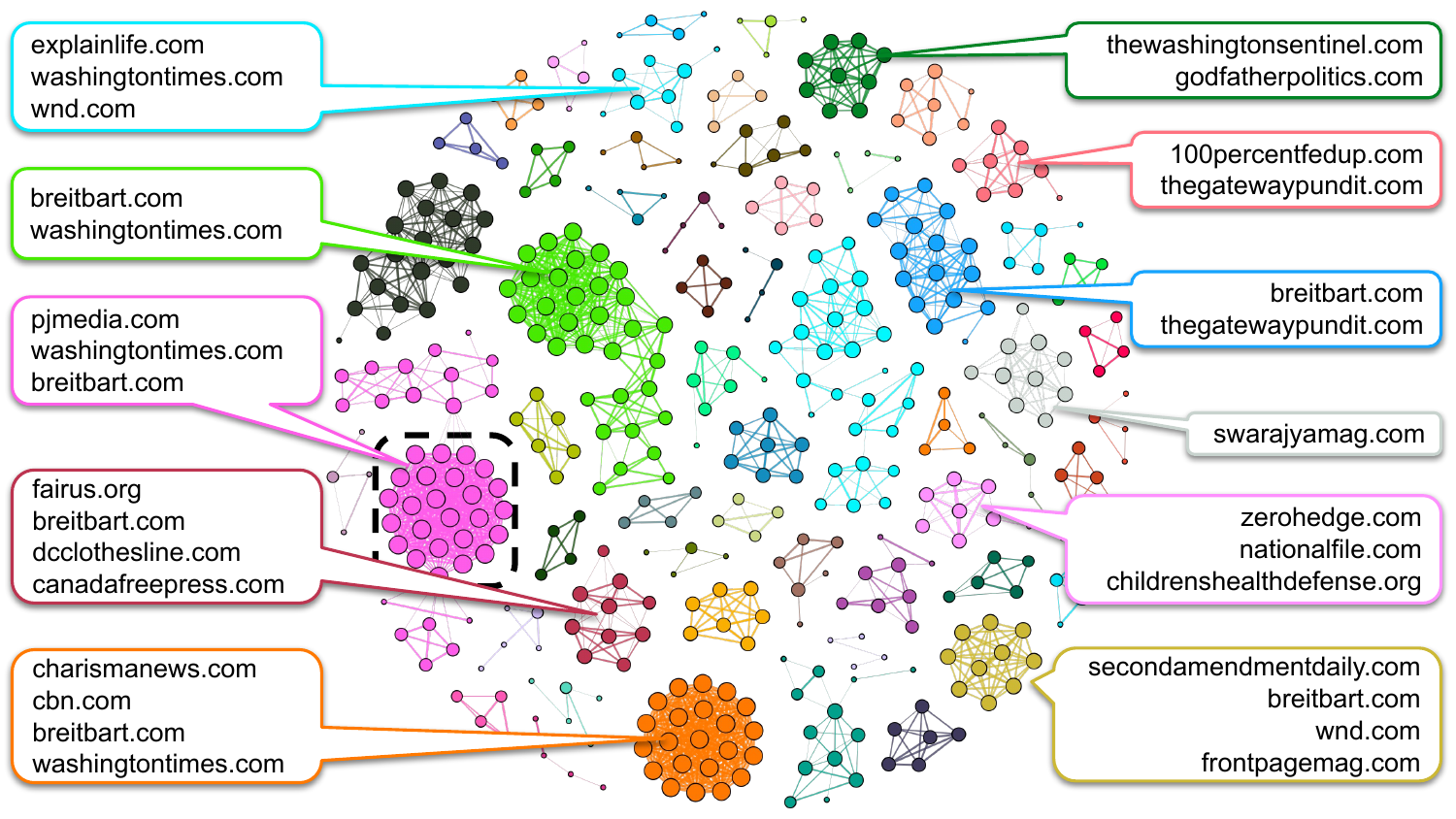}
    \caption{Networks showing clusters that share suspiciously similar sets of sources on \textbf{(top)} Twitter and \textbf{(bottom)} Facebook. Nodes represent Twitter users or Facebook pages/groups. The size of the each node is proportional to its degree. Edges are drawn between pairs of nodes that share an unlikely high number of the same low-credibility domains. The edge weight represents the number of co-shared domains. The most shared sources are annotated for some of the clusters. Facebook pages associated with Salem Media Group radio stations are highlighted by a dashed box.}
    \label{fig:co_url_network}
\end{figure*}

Fig.~\ref{fig:co_url_network} shows densely connected components in the co-domain networks for  Twitter and Facebook.
These clusters of accounts share suspiciously similar sets of sources. They likely act in a coordinated fashion to amplify Infodemic messages, and are possibly controlled by the same entity or organization. 
We highlight the fact that using a more stringent threshold on the Twitter dataset yields a higher number of clusters than a more lax threshold on the Facebook dataset. However, this does not necessarily imply a higher level of abuse on Twitter; it could be due to the difference in the units of analysis. On Facebook, we only have access to public groups and pages with a bias toward high popularity, and not to all accounts as on Twitter.   

An examination of the sources shared by the suspicious clusters on both platforms shows that they are predominantly right-leaning and mostly U.S.-centric.
The list of domains shared by likely coordinated accounts on Twitter is mostly concentrated on the leading low-credibility sources, such as \texttt{breitbart.com} and \texttt{thegatewaypundit.com}, while likely coordinated groups and pages on Facebook link to a more varied list of sources.
Some of the amplified websites feature polarized rhetoric, such as defending against ``attack by enemies'' (see \texttt{www.frontpagemag.com/about}) or accusations of ``liberal bias'' (\texttt{cnsnews.com/about-us}), among others. 
Additionally, there are clusters on both platforms that share Russian state-affiliated media such as \texttt{rt.com} and an Indian right-wing magazine (\texttt{swarajyamag.com}).

In terms of the composition of the clusters, they mostly consist of users, pages, and groups that mention President Trump or his campaign slogans. 
Some of the Facebook clusters are notable because they consist of groups or pages that are owned by organizations with a wider reach beyond the platform, or that are given an appearance of credibility by being verified.
Examples of the former are the pages associated with \textit{The Answer} radio stations (highlighted in Fig.~\ref{fig:co_url_network}). These are among 100 stations owned by the publicly traded Salem Media Group, which also airs content on 3,100 affiliate stations. 
Examples of verified pages engaged in likely coordinated behavior are those affiliated with the non-profit Media Research Center, some of which have millions of followers. 
On Twitter, some of the clusters include accounts with tens of thousands of followers. Many of the suspicious accounts in Fig.~\ref{fig:co_url_network} no longer exist.  
 
\subsection{Role of social bots}

We are interested in revealing the role of inauthentic actors in spreading low-credibility information on social media.
One type of inauthentic behavior stems from accounts controlled in part by algorithms, known as  \textit{social bots}~\cite{ferrara2016rise}.
Malicious bots are known to spread low-credibility information~\cite{shao2018spread} and in particular create confusion in the online debate around health-related topics like vaccination~\cite{broniatowski2018weaponized}.  

We adopt BotometerLite (\url{rapidapi.com/OSoMe/api/botometer-pro}), a publicly-available tool that allows efficient bot detection on Twitter~\cite{yang2020scalable}. 
BotometerLite generates a bot score between 0 and 1 for each Twitter account; higher scores indicate bot-like profiles.
To the best of our knowledge, there are no similar techniques designed for Facebook because insufficient training data is available. Therefore we limit this analysis to Twitter.

\begin{figure}
    \centering
    \includegraphics[width=0.6\columnwidth]{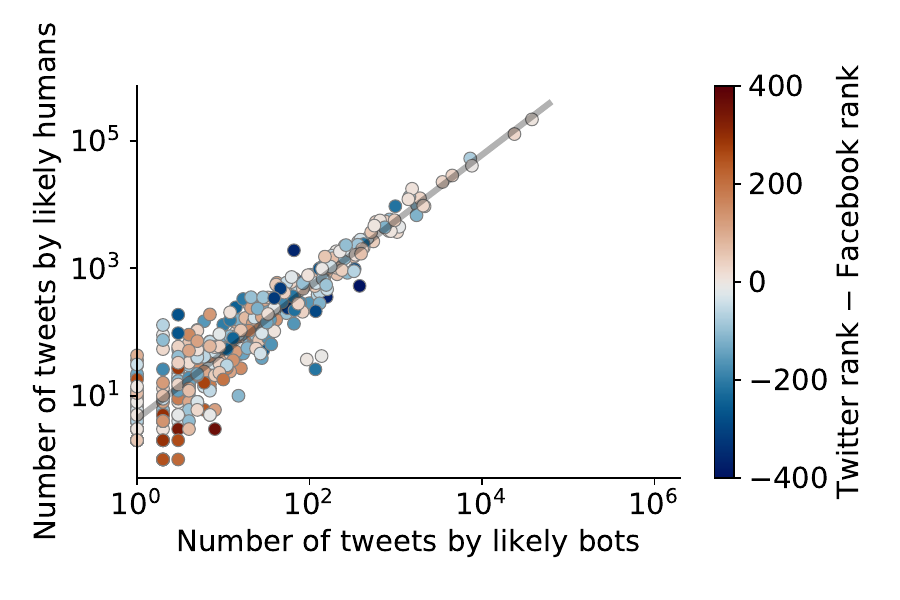}
    \caption{
    Total number of tweets with links posted by likely humans vs.~likely bots for each low-credibility source.
    The slope of the fitted line is 1.04.
    The color of each source represents the difference between its popularity rank on the two platforms. Red means more popular on Facebook, blue more popular on Twitter.
    }
    \label{fig:bot_rate}
\end{figure}

When applying BotometerLite to our Twitter dataset, we use 0.5 as the threshold to categorize accounts as likely humans or likely bots.
For each domain, we calculate the total number of original tweets plus retweets authored by likely humans ($n_h$) and bots ($n_b$).
We plot the relationship between the two in Fig.~\ref{fig:bot_rate}. The linear trend on the log-log plot signifies a power law $n_h \sim n_b^\gamma$ with exponent $\gamma \approx 1.04$, suggesting a weak level of bot amplification (4\%)~\cite{shao2018spread}. 

While we are unable to perform automation detection on Facebook groups and pages, the ranks of the low-credibility sources on both platforms allow us to investigate whether sources with more Twitter bot activity are more prevalent on Twitter or Facebook.
For each domain, we calculate the difference of its ranks on Twitter and Facebook and use the value of the difference to color the dots in Fig.~\ref{fig:bot_rate}.
The results show that sources with more bot activity on Twitter are equally shared on both platforms. 

\section{Discussion}

In this paper, we provide the first comparison between the prevalence of low-credibility content related to the COVID-19 pandemic on two major social media platforms, namely Twitter and Facebook. Our results indicate that the primary drivers of low-credibility information tend to be high-profile, official, and verified accounts. We also find evidence of coordination among accounts spreading Infodemic content on both platforms, including many controlled by influential organizations. Since automated accounts do not appear to play a strong role in amplifying content, these results indicate that the COVID-19 Infodemic is an overt, rather than a covert, phenomenon.

We find that low-credibility content, as a whole, has higher prevalence than content from any single high-credibility source. However, there is evidence of differences in the misinformation ecosystems of the two platforms, with many low-credibility websites and suspicious YouTube videos at higher prevalence on one platform when compared to the other.
Such a discrepancy might be due to a combination of the supply and demand factors.
On the supply side, the official accounts associated with specific low-credibility websites are not symmetrically present on both platforms.
On the demand side, the two platforms have very different user demographics. 
According to recent surveys, 69\% of adults in the U.S. say they use Facebook, but only 22\% of adults are on Twitter.
Further, while Facebook usage is relatively common across a range of demographic groups, Twitter users tend to be younger, more educated, and have higher than average income.
Finally, Facebook is a pathway to consuming online news for around 43\% of U.S. adults, while the same number for Twitter is 12\%~\cite{perrin2019share,pew2019b}.

During the first months of the pandemic, we observe similar surges of low-credibility content on both platforms. 
The strong correlation between the timelines of low- and high-credibility content volume reveals that these peaks were likely driven by public attention to the crisis rather than by bursts of malicious content. 

Our results provide us with a way to assess how effective the two platforms have been at combating the Infodemic. The ratio of low- to high-credibility information on Facebook is lower than on Twitter, suggesting that  Facebook may be more effective. On the other hand, we also find that verified accounts played a stronger role on Facebook than Twitter in spreading low-credibility content. However, the accuracy of these comparisons is subject to the different data collection biases. 
Suspicious YouTube uploads also exhibit an asymmetric prevalence between Facebook and Twitter. As stated previously, this may be partly a result of uploaders recycling sections of videos and uploading the content with a new video ID. Having such duplicates can mean that one version becomes popular on Facebook and another on Twitter, each potentially shared by a different demographic. This asymmetry might also be driven by Twitter users being more likely to flag videos. YouTube may then quickly remove reported videos before Facebook users have a chance to share them.

There are a number of limitations to our work. As we have remarked throughout the paper, differences between platform data availability and biases in sampling and selection make direct and fair comparisons impossible in many cases. 
The content collected from the Twitter Decahose is biased toward active users due to being sampled on a per-tweet basis. 
The Facebook accounts provided by CrowdTangle are biased toward popular pages and public groups, and data availability is also based upon requests made by other researchers.
The small set of keywords driving our data collection pipeline may have introduced additional biases in the analyses. This is an inevitable limitation of any collection system, including the Twitter COVID-19 stream (\url{developer.twitter.com/en/docs/labs/covid19-stream/filtering-rules}). 
The use of source-level rather than article-level labels for selecting low-credibility content is necessary~\cite{Lazer-fake-news-2018}, but not ideal; some links from low-credibility sources may point to  credible information. 
In addition, the list of low-credibility sources was not specifically tailored to our subject of inquiry. 
Finally, we do not have access to many deleted Twitter and Facebook posts, which may lead to an underestimation of the Infodemic's prevalence. All of these limitations highlight the need for cross-platform, privacy-sensitive protocols for sharing data with researchers~\cite{HKS-MR}.

Low-credibility information on the pandemic is an ongoing concern for society.
Our study raises a number of questions.
For example, user demographics might strongly affect the consumption of low-credibility information on social media: how do users in distinct demographic groups interact with different information sources?
The answer to this question can lead to a better understanding of the Infodemic and more effective moderation strategies, but will require methods that scale with the nature of big data from social media. 
Another critical question is how social media platforms are handling the flow of information and allowing dangerous content to spread.
Regrettably, since we find that high-status accounts play an important role, addressing this problem will prove difficult.
As Twitter and Facebook have increased their moderation of COVID-19 misinformation, they have been accused of political bias.
While there are many legal and ethical considerations around free speech and censorship, our work suggests that these questions cannot be avoided and are an important part of the debate around how we can improve our information ecosystem.


\bibliographystyle{unsrt}
\bibliography{ref.bib}

\begin{thebibliography}{10}

\bibitem{lyu_community_2020}
Wei Lyu and George~L. Wehby.
\newblock Community use of face masks and {COVID}-19: Evidence from a natural
  experiment of state mandates in the {US}.
\newblock {\em Health Affairs}, 39(8):1419--1425, June 2020.

\bibitem{loomba2021measuring}
Sahil Loomba, Alexandre de~Figueiredo, Simon~J Piatek, Kristen de~Graaf, and
  Heidi~J Larson.
\newblock Measuring the impact of {COVID-19} vaccine misinformation on
  vaccination intent in the {UK} and {USA}.
\newblock {\em Nature Human Behaviour}, 2021.

\bibitem{mitchell_americans_2020}
Amy Mitchell and J.~Baxter Oliphant.
\newblock Americans immersed in coronavirus news; most think media are doing
  fairly well covering it.
\newblock Pew Research Center, March 2020.
\newblock \url{https://pewrsr.ch/3dbTpxs} (Accessed November 2020).

\bibitem{schaeffer_nearly_2020}
Katherine Schaeffer.
\newblock Nearly three-in-ten {Americans} believe {COVID}-19 was made in a lab.
\newblock Pew Research Center, April 2020.
\newblock \url{https://pewrsr.ch/2XlJqAa} (Accessed November 2020).

\bibitem{nightingale_quantifying_2020}
Sophie Nightingale, Marc Faddoul, and Hany Farid.
\newblock Quantifying the reach and belief in {COVID}-19 misinformation.
\newblock {\em arXiv:2006.08830}, June 2020.

\bibitem{roozenbeek_susceptibility_2020}
Jon Roozenbeek, Claudia~R. Schneider, Sarah Dryhurst, John Kerr, Alexandra
  L.~J. Freeman, Gabriel Recchia, Anne~Marthe van~der Bles, and Sander van~der
  Linden.
\newblock Susceptibility to misinformation about {COVID}-19 around the world.
\newblock {\em Royal Society Open Science}, 7(10):201199, 2020.
\newblock Publisher: Royal Society.

\bibitem{zarocostas2020fight}
John Zarocostas.
\newblock How to fight an infodemic.
\newblock {\em The Lancet}, 395(10225):676, 2020.

\bibitem{vosoughi2018spread}
Soroush Vosoughi, Deb Roy, and Sinan Aral.
\newblock The spread of true and false news online.
\newblock {\em Science}, 359(6380):1146--1151, 2018.

\bibitem{shao2018spread}
Chengcheng Shao, Giovanni~Luca Ciampaglia, Onur Varol, Kai-Cheng Yang,
  Alessandro Flammini, and Filippo Menczer.
\newblock The spread of low-credibility content by social bots.
\newblock {\em Nature Communications}, 9:4787, 2018.

\bibitem{stella2018bots}
Massimo Stella, Emilio Ferrara, and Manlio De~Domenico.
\newblock Bots increase exposure to negative and inflammatory content in online
  social systems.
\newblock {\em Proceedings of the National Academy of Sciences},
  115(49):12435--12440, 2018.

\bibitem{ferrara_misinformation_2020}
Emilio Ferrara, Stefano Cresci, and Luca Luceri.
\newblock Misinformation, manipulation, and abuse on social media in the era of
  {COVID}-19.
\newblock {\em Journal of Computational Social Science}, 3(2):271--277,
  November 2020.

\bibitem{allcott_trends_2019}
Hunt Allcott, Matthew Gentzkow, and Chuan Yu.
\newblock Trends in the diffusion of misinformation on social media.
\newblock {\em Research \& Politics}, 6(2):2053168019848554, April 2019.

\bibitem{ferrara_what_2020}
Emilio Ferrara.
\newblock What types of {COVID}-19 conspiracies are populated by {Twitter}
  bots?
\newblock {\em First Monday}, 25(6), 2020.

\bibitem{pacheco2020uncovering}
Diogo Pacheco, Pik-Mai Hui, Christopher Torres-Lugo, Bao~Tran Truong,
  Alessandro Flammini, and Filippo Menczer.
\newblock Uncovering coordinated networks on social media: Methods and case
  studies.
\newblock In {\em Proceedings of the AAAI International Conference on Web and
  Social Media (ICWSM)}, 2021.
\newblock Forthcoming.

\bibitem{Lazer-fake-news-2018}
David Lazer, Matthew Baum, Yochai Benkler, Adam Berinsky, Kelly Greenhill,
  et~al.
\newblock The science of fake news.
\newblock {\em Science}, 359(6380):1094--1096, 2018.

\bibitem{eysenbach2002empirical}
Gunther Eysenbach, John Powell, Oliver Kuss, and Eun-Ryoung Sa.
\newblock Empirical studies assessing the quality of health information for
  consumers on the world wide web: a systematic review.
\newblock {\em JAMA}, 287(20):2691--2700, 2002.

\bibitem{eysenbach2002infodemiology}
Gunther Eysenbach.
\newblock Infodemiology: The epidemiology of (mis) information.
\newblock {\em The American Journal of Medicine}, 113(9):763--765, 2002.

\bibitem{jin2014misinformation}
Fang Jin, Wei Wang, Liang Zhao, Edward Dougherty, Yang Cao, C~Lu, and Naren
  Ramakrishnan.
\newblock Misinformation propagation in the age of {Twitter}.
\newblock {\em IEEE Annals of the History of Computing}, 47(12):90--94, 2014.

\bibitem{pathak2015youtube}
Ranjan Pathak, Dilli~Ram Poudel, Paras Karmacharya, Amrit Pathak, Madan~Raj
  Aryal, Maryam Mahmood, and Anthony~A Donato.
\newblock {YouTube as a source of information on Ebola virus disease}.
\newblock {\em North American Journal of Medical Sciences}, 7(7):306, 2015.

\bibitem{fung2016social}
Isaac Chun-Hai Fung, King-Wa Fu, Chung-Hong Chan, Benedict Shing~Bun Chan,
  Chi-Ngai Cheung, Thomas Abraham, and Zion Tsz~Ho Tse.
\newblock Social media's initial reaction to information and misinformation on
  {Ebola}, {August} 2014: facts and rumors.
\newblock {\em Public Health Reports}, 131(3):461--473, 2016.

\bibitem{sell2020misinformation}
Tara~Kirk Sell, Divya Hosangadi, and Marc Trotochaud.
\newblock Misinformation and the us ebola communication crisis: analyzing the
  veracity and content of social media messages related to a fear-inducing
  infectious disease outbreak.
\newblock {\em BMC Public Health}, 20:1--10, 2020.

\bibitem{seltzer2017public}
EK~Seltzer, E~Horst-Martz, M~Lu, and RM~Merchant.
\newblock Public sentiment and discourse about {Zika} virus on {Instagram}.
\newblock {\em Public Health}, 150:170--175, 2017.

\bibitem{sharma2017zika}
Megha Sharma, Kapil Yadav, Nitika Yadav, and Keith~C Ferdinand.
\newblock Zika virus pandemic—analysis of {Facebook} as a social media health
  information platform.
\newblock {\em American Journal of Infection Control}, 45(3):301--302, 2017.

\bibitem{bora2018internet}
Kaustubh Bora, Dulmoni Das, Bhupen Barman, and Probodh Borah.
\newblock {Are internet videos useful sources of information during global
  public health emergencies? A case study of YouTube videos during the 2015--16
  Zika virus pandemic}.
\newblock {\em Pathogens and Global Health}, 112(6):320--328, 2018.

\bibitem{wood2018propagating}
Michael~J Wood.
\newblock {Propagating and debunking conspiracy theories on Twitter during the
  2015--2016 Zika virus outbreak}.
\newblock {\em Cyberpsychology, Behavior, and Social Networking},
  21(8):485--490, 2018.

\bibitem{mahoney2015digital}
L~Meghan Mahoney, Tang Tang, Kai Ji, and Jessica Ulrich-Schad.
\newblock The digital distribution of public health news surrounding the human
  papillomavirus vaccination: a longitudinal infodemiology study.
\newblock {\em JMIR Public Health and Surveillance}, 1(1):e2, 2015.

\bibitem{bahk2016publicly}
Chi~Y Bahk, Melissa Cumming, Louisa Paushter, Lawrence~C Madoff, Angus Thomson,
  and John~S Brownstein.
\newblock Publicly available online tool facilitates real-time monitoring of
  vaccine conversations and sentiments.
\newblock {\em Health affairs}, 35(2):341--347, 2016.

\bibitem{donzelli2018misinformation}
Gabriele Donzelli, Giacomo Palomba, Ileana Federigi, Francesco Aquino, Lorenzo
  Cioni, Marco Verani, Annalaura Carducci, and Pierluigi Lopalco.
\newblock Misinformation on vaccination: A quantitative analysis of youtube
  videos.
\newblock {\em Human Vaccines \& Immunotherapeutics}, 14(7):1654--1659, 2018.

\bibitem{panatto2018comprehensive}
Donatella Panatto, Daniela Amicizia, Lucia Arata, Piero~Luigi Lai, and Roberto
  Gasparini.
\newblock A comprehensive analysis of {Italian} web pages mentioning
  squalene-based influenza vaccine adjuvants reveals a high prevalence of
  misinformation.
\newblock {\em Human Vaccines \& Immunotherapeutics}, 14(4):969--977, 2018.

\bibitem{schmidt2018polarization}
Ana~Luc{\'\i}a Schmidt, Fabiana Zollo, Antonio Scala, Cornelia Betsch, and
  Walter Quattrociocchi.
\newblock Polarization of the vaccination debate on {Facebook}.
\newblock {\em Vaccine}, 36(25):3606--3612, 2018.

\bibitem{deverna2021covaxxy}
Matthew~R. DeVerna, Francesco Pierri, Bao~Tran Truong, John Bollenbacher, David
  Axelrod, Niklas Loynes, Christopher Torres-Lugo, Kai-Cheng Yang, Filippo
  Menczer, and John Bryden.
\newblock {CoVaxxy: A global collection of English Twitter posts about COVID-19
  vaccines}.
\newblock {\em arXiv preprint arXiv:2101.07694}, 2021.

\bibitem{pulido_covid-19_2020}
Cristina~M Pulido, Beatriz Villarejo-Carballido, Gisela Redondo-Sama, and Aitor
  Gómez.
\newblock {COVID}-19 infodemic: {More} retweets for science-based information
  on coronavirus than for false information.
\newblock {\em International Sociology}, 35(4):377--392, July 2020.

\bibitem{al-rakhami_lies_2020}
M.~S. Al-Rakhami and A.~M. Al-Amri.
\newblock Lies {Kill}, {Facts} {Save}: {Detecting} {COVID}-19 {Misinformation}
  in {Twitter}.
\newblock {\em IEEE Access}, 8:155961--155970, 2020.

\bibitem{memon_characterizing_2020}
Shahan~Ali Memon and Kathleen~M. Carley.
\newblock Characterizing {COVID}-19 misinformation communities using a novel
  {Twitter} dataset.
\newblock {\em arXiv:2008.00791}, September 2020.

\bibitem{kouzy2020coronavirus}
Ramez Kouzy, Joseph Abi~Jaoude, Afif Kraitem, Molly~B El~Alam, Basil Karam,
  Elio Adib, Jabra Zarka, Cindy Traboulsi, Elie~W Akl, and Khalil Baddour.
\newblock Coronavirus goes viral: quantifying the {COVID-19} misinformation
  epidemic on {Twitter}.
\newblock {\em Cureus}, 12(3), 2020.

\bibitem{li2020youtube}
Heidi Oi-Yee Li, Adrian Bailey, David Huynh, and James Chan.
\newblock {YouTube} as a source of information on {COVID}-19: a pandemic of
  misinformation?
\newblock {\em BMJ Global Health}, 5(5):e002604, 2020.

\bibitem{Dutta_YTcovid_2020}
A.~Dutta, N.~Beriwal, L.~M. Van~Breugel, et~al.
\newblock {YouTube} as a source of medical and epidemiological information
  during {COVID}-19 pandemic: A cross-sectional study of content across six
  languages around the globe.
\newblock {\em Cureus}, 12(6):e8622, 2020.

\bibitem{broniatowski_covid-19_2020}
David~A. Broniatowski, Daniel Kerchner, Fouzia Farooq, Xiaolei Huang, Amelia~M.
  Jamison, Mark Dredze, and Sandra~Crouse Quinn.
\newblock The {COVID}-19 social media {Infodemic} reflects uncertainty and
  state-sponsored propaganda.
\newblock {\em arXiv:2007.09682}, July 2020.

\bibitem{cinelli_covid-19_2020}
Matteo Cinelli, Walter Quattrociocchi, Alessandro Galeazzi, Carlo~Michele
  Valensise, Emanuele Brugnoli, Ana~Lucia Schmidt, Paola Zola, Fabiana Zollo,
  and Antonio Scala.
\newblock The {COVID}-19 social media {Infodemic}.
\newblock {\em Scientific Reports}, 10(1):16598, 2020.

\bibitem{gallotti2020assessing}
Riccardo Gallotti, Francesco Valle, Nicola Castaldo, Pierluigi Sacco, and
  Manlio De~Domenico.
\newblock {Assessing the risks of `infodemics' in response to COVID-19
  epidemics}.
\newblock {\em Nature Human Behaviour}, 4:1285–1293, 2020.

\bibitem{singh_first_2020}
Lisa Singh, Shweta Bansal, Leticia Bode, Ceren Budak, Guangqing Chi, Kornraphop
  Kawintiranon, Colton Padden, Rebecca Vanarsdall, Emily Vraga, and Yanchen
  Wang.
\newblock A first look at {COVID}-19 information and misinformation sharing on
  {Twitter}.
\newblock {\em arXiv:2003.13907}, March 2020.

\bibitem{yang2020prevalence}
Kai-Cheng Yang, Christopher Torres-Lugo, and Filippo Menczer.
\newblock Prevalence of low-credibility information on twitter during the
  {COVID}-19 outbreak.
\newblock In {\em Proceedings of the ICWSM International Workshop on Cyber
  Social Threats}, 2020.

\bibitem{singh2020understanding}
Lisa Singh, Leticia Bode, Ceren Budak, Kornraphop Kawintiranon, Colton Padden,
  and Emily Vraga.
\newblock {Understanding high-and low-quality URL Sharing on COVID-19 Twitter
  streams}.
\newblock {\em Journal of Computational Social Science}, 3:343–366, 2020.

\bibitem{guarino2021information}
Stefano Guarino, Francesco Pierri, Marco Di~Giovanni, and Alessandro Celestini.
\newblock Information disorders during the {COVID}-19 infodemic: The case of
  {Italian} {Facebook}.
\newblock {\em Online Social Networks and Media}, 22:100124, 2021.

\bibitem{chou2018addressing}
Wen-Ying~Sylvia Chou, April Oh, and William~MP Klein.
\newblock Addressing health-related misinformation on social media.
\newblock {\em JAMA}, 320(23):2417--2418, 2018.

\bibitem{grinberg2019fake}
Nir Grinberg, Kenneth Joseph, Lisa Friedland, Briony Swire-Thompson, and David
  Lazer.
\newblock {Fake news on Twitter during the 2016 US presidential election}.
\newblock {\em Science}, 363(6425):374--378, 2019.

\bibitem{pennycook2019fighting}
Gordon Pennycook and David~G Rand.
\newblock Fighting misinformation on social media using crowdsourced judgments
  of news source quality.
\newblock {\em Proceedings of the National Academy of Sciences},
  116(7):2521--2526, 2019.

\bibitem{bovet2019influence}
Alexandre Bovet and Hern{\'a}n~A Makse.
\newblock {Influence of fake news in Twitter during the 2016 US presidential
  election}.
\newblock {\em Nature Communications}, 10(1):1--14, 2019.

\bibitem{dataset}
Kai-Cheng Yang, Francesco Pierri, Pik-Mai Hui, David Axelrod, Christopher
  Torres-Lugo, John Bryden, and Filippo Menczer.
\newblock {Dataset for paper: The COVID-19 Infodemic: Twitter versus Facebook}.
\newblock Zenodo, 2020.
\newblock \url{https://doi.org/10.5281/zenodo.4313903}.

\bibitem{pew2014}
Amy Mitchell, Jeffrey Gottfried, Jocelyn Kiley, and Katerina~Eva Matsa.
\newblock Political polarization \& media habits.
\newblock Pew Research Center, 2014.
\newblock \url{http://pewrsr.ch/1vZ9MnM} (Accessed November 2020).

\bibitem{Pierri2020epj}
Francesco Pierri, Carlo Piccardi, and Stefano Ceri.
\newblock A multi-layer approach to disinformation detection in {US} and
  {Italian} news spreading on twitter.
\newblock {\em EPJ Data Science}, 9(35), 2020.

\bibitem{Pierri2020scirep}
Francesco Pierri, Carlo Piccardi, and Stefano Ceri.
\newblock Topology comparison of {Twitter} diffusion networks effectively
  reveals misleading news.
\newblock {\em Scientific Reports}, 10:1372, 2020.

\bibitem{davis2016osome}
Clayton~A Davis, Giovanni~Luca Ciampaglia, Luca~Maria Aiello, Keychul Chung,
  Michael~D Conover, Emilio Ferrara, Alessandro Flammini, et~al.
\newblock {OSoMe: the IUNI observatory on social media}.
\newblock {\em PeerJ Computer Science}, 2:e87, 2016.

\bibitem{crowdtangle}
{CrowdTangle Team}.
\newblock {CrowdTangle. Menlo Park, CA: Facebook.}, 2020.
\newblock Accessed November 2020.

\bibitem{knuutila2020covid}
Aleksi Knuutila, Aliaksandr Herasimenka, Hubert Au, Jonathan Bright, Rasmus
  Nielsen, and Philip~N Howard.
\newblock {COVID-related misinformation on YouTube: The spread of
  misinformation videos on social media and the effectiveness of platform
  policies}.
\newblock Oxford, UK: Project on Computational Propaganda, 2020.
\newblock
  \url{https://comprop.oii.ox.ac.uk/research/posts/youtube-platform-policies/}.

\bibitem{wilson2020cross}
Tom Wilson and Kate Starbird.
\newblock Cross-platform disinformation campaigns: Lessons learned and next
  steps.
\newblock {\em Harvard Kennedy School Misinformation Review}, 1(1), 2020.

\bibitem{Goobie_idiopathic_2019}
Gillian~C. Goobie, Sabina~A. Gulera, Kerri~A. Johannson, Jolene~H. Fisher, and
  Christopher~J. Ryerson.
\newblock {YouTube} videos as a source of misinformation on idiopathic
  pulmonary fibrosis.
\newblock {\em Annals of the American Thoracic Society}, 16(5):572–--579,
  2019.

\bibitem{Nikolov2018biases}
Dimitar Nikolov, Mounia Lalmas, Alessandro Flammini, and Filippo Menczer.
\newblock Quantifying biases in online information exposure.
\newblock {\em Journal of the Association for Information Science and
  Technology}, 70(3):218--229, 2019.

\bibitem{nizzoli2020coordinated}
Leonardo Nizzoli, Serena Tardelli, Marco Avvenuti, Stefano Cresci, and Maurizio
  Tesconi.
\newblock Coordinated behavior on social media in 2019 {UK} general election.
\newblock {\em arXiv:2008.08370}, 2020.

\bibitem{sharma2020identifying}
Karishma Sharma, Emilio Ferrara, and Yan Liu.
\newblock Identifying coordinated accounts in disinformation campaigns.
\newblock {\em arXiv:2008.11308}, 2020.

\bibitem{jones1972statistical}
Karen Sp\"arck~Jones.
\newblock A statistical interpretation of term specificity and its application
  in retrieval.
\newblock {\em Journal of Documentation}, 28(1):11--21, 1972.

\bibitem{ferrara2016rise}
Emilio Ferrara, Onur Varol, Clayton Davis, Filippo Menczer, and Alessandro
  Flammini.
\newblock The rise of social bots.
\newblock {\em Communications of the ACM}, 59(7):96--104, 2016.

\bibitem{broniatowski2018weaponized}
David~A Broniatowski, Amelia~M Jamison, SiHua Qi, Lulwah AlKulaib, Tao Chen,
  Adrian Benton, Sandra~C Quinn, and Mark Dredze.
\newblock {Weaponized health communication: Twitter bots and Russian trolls
  amplify the vaccine debate}.
\newblock {\em American Journal of Public Health}, 108(10):1378--1384, 2018.

\bibitem{yang2020scalable}
Kai-Cheng Yang, Onur Varol, Pik-Mai Hui, and Filippo Menczer.
\newblock Scalable and generalizable social bot detection through data
  selection.
\newblock {\em Proceedings of the AAAI Conference on Artificial Intelligence},
  34(1):1096--1103, 2020.

\bibitem{perrin2019share}
Andrew Perrin and Monica Anderson.
\newblock {Share of US adults using social media, including Facebook, is mostly
  unchanged since 2018}.
\newblock Pew Research Center, 2019.
\newblock \url{https://pewrsr.ch/2VxJuJ3} (Accessed February 2021).

\bibitem{pew2019b}
Stephan Wojcik and Hughes Adam.
\newblock Sizing up {Twitter} users.
\newblock Pew Research Center, 2019.
\newblock \url{https://pewrsr.ch/2VUkzj4} (Accessed February 2021).

\bibitem{HKS-MR}
Irene~V. Pasquetto, Briony Swire-Thompson, et~al.
\newblock Tackling misinformation: What researchers could do with social media
  data.
\newblock {\em Harvard Kennedy School Misinformation Review}, 1(8), 2020.

\end{thebibliography}

\end{document}